%% file: main.tex
\documentclass[journal]{IEEEtran}
\usepackage{cite}

\ifCLASSINFOpdf
\else
\fi
\usepackage[dvipsnames]{xcolor}
\usepackage{wrapfig}
\usepackage[acronym,nomain]{glossaries}
\usepackage{amsmath,amssymb}
\usepackage{xfrac}
\usepackage{subcaption}  
\usepackage{tikz,pgfplots}
\usepackage{floatrow}
\usepackage{scalerel}
\usepackage{dblfloatfix} 
\usepackage[many]{tcolorbox} 			
\usepackage{colortbl}
\usetikzlibrary{backgrounds,shapes,arrows,positioning,calc}
\usepackage[pdftex,colorlinks=true,bookmarks=false,citecolor=blue,urlcolor=blue]{hyperref} 
\pgfplotsset{compat=newest}
\pgfplotsset{every axis/.append style={
                    label style={font=\large},
                    tick label style={font=\large},
                    axis label style={font=\large},
                    scaled y ticks=false}
            every tick label/.append style={font=\large}}                    
\pgfdeclarelayer{background}
\pgfdeclarelayer{foreground}
\pgfsetlayers{background,main,foreground}
\newfloatcommand{capbtabbox}{table}[][0.9\FBwidth]
\usepackage[pdftex,colorlinks=true,bookmarks=false,citecolor=blue,urlcolor=blue]{hyperref} 

\definecolor{myblue}{rgb}{0.00000,0.44700,0.74100}%
\definecolor{myred}{rgb}{1.0, 0.01, 0.24}
\input{acronyms.tex}

\renewcommand{\thesection}{\Roman{section}}
\renewcommand{\thesubsection}{\thesection.\Roman{subsection}}

\begin{document}

\author{Gabriele~Liga,~\IEEEmembership{Member,~IEEE,}
        Bin~Chen,~\IEEEmembership{Member,~IEEE,}
        Sjoerd~van~der~Heide,~\IEEEmembership{Student Member,~IEEE,}
        Alireza~Sheikh,~\IEEEmembership{Member,~IEEE,}
        Menno van der Hout,
        Chigo~Okonkwo,~\IEEEmembership{Senior Member,~IEEE,}
        and~Alex~Alvarado,~\IEEEmembership{Senior Member,~IEEE}
\thanks{G. Liga, A. Sheikh, and A. Alvarado are with the Signal Processing Systems (SPS) Group, Department of Electrical Engineering, Eindhoven  University  of  Technology,  The  Netherlands. S.~van~der~Heide, M. van~der~Hout and C. Okonkwo are with the Electro-Optical  Communications (ECO) Group,
Department of Electrical Engineering, Eindhoven University of Technology, The Netherlands \mbox{e-mail:} g.liga@tue.nl.}
\thanks{B. Chen is with the School of Computer Science and Information Engineering, Hefei University of Technology, China,  and  the  Signal  Processing  Systems  (SPS)  Group,  Department  of  Electrical  Engineering,  Eindhoven  University  of Technology, The Netherlands.}}

\title{30\% Reach Increase via Low-complexity Hybrid HD/SD FEC and Nonlinearity-tolerant 4D Modulation}

\maketitle


\begin{abstract}
Current optical coherent transponders technology is driving data rates towards 1 Tb/s/$\lambda$ and beyond. This trend requires both high-performance coded modulation schemes and efficient implementation of the forward-error-correction (FEC) decoder. A possible solution to this problem is combining advanced multidimensional modulation formats with low-complexity \emph{hybrid} HD/SD FEC decoders. Following this rationale, in this paper we combine two recently introduced coded modulation techniques: the geometrically-shaped 4D-64 polarization ring-switched and the soft-aided bit-marking-scaled reliability decoder. This joint scheme enabled us to experimentally demonstrate the transmission of 11$\times$×218 Gbit/s channels over transatlantic distances at 5.2 bit/4D-sym. Furthermore, a 30\% reach increase is demonstrated over PM-8QAM and conventional HD-FEC decoding for product codes.
\end{abstract}


\section{Introduction}
Commercially available optical line rates have today reached 800 Gbit/s/$\lambda$, and they are rapidly heading towards 1 Tbit/s/$\lambda$ and beyond, thus exerting unprecedented pressure on the optical transponder electronics. Such high data rates require high-performance yet implementation-efficient coded modulation schemes. Whilst high spectral-efficiency transmission can be provided by cleverly designed high-order modulation formats, low-complexity \gls{FEC} schemes are key to keep manageable power consumption in next-generation high-speed optical line cards.


Although \gls{SD}-FEC decoding represents the current gold-standard for \gls{FEC} decoders in long-haul coherent optical communications, it entails high decoding complexity and dataflow, which makes its adoption into next-generation high-throughput line cards challenging \cite{Smith2012}. Therefore, solutions which trade-off performance for a lower decoding complexity are becoming increasingly attractive \cite{OIF400G}. Along the path traced by Chase in 1972 \cite{Chase1972}, hybrid \gls{HD}/SD decoders have been recently reproposed in optical communications as a low-complexity alternative to fully-fledged SD-FEC schemes \cite{Yi2019,Sheikh2019tcom,FougstedtOFC2019,Liga2019ECOC}. In these schemes, reliability metrics are used to assist a standard HD decoder to improve its performance, whilst keeping the complexity of the overall decoder of the same order as that of algebraic HD decoding.
These new decoding algorithms have been applied to both \gls{PCs} and staircase codes, showing substantial coding gains (0.2--0.8 dB) compared to its traditional HD counterpart, referred to as \gls{iBDD} \cite[Sec.~II-A]{She18b}. One such decoding algorithm is the \gls{SABM} algorithm which was introduced in \cite{YiISTC2018} and later extended in \cite{Liga2019ECOC} to incorporate so-called scaled reliabilities (SRs), defined in \cite{Sheikh2019tcom}, in the decoding process. This variant of \gls{SABM}, named SABM-SR, was shown to outperform iBDD and \gls{SABM} by up to 0.8 dB and 0.3 dB, respectively, with only minor additional complexity \cite{Liga2019ECOC}. 
\begin{figure*}[!t]
\begin{minipage}{0.23\columnwidth}
\vspace{-0.5cm}
\captionsetup[subfigure]{oneside,margin={1.2cm,0cm},skip=-1pt}
\begin{subfigure}[b]{\columnwidth}
\begin{equation*}
n\left\{\left(
\begin{array}{cc>{\columncolor{blue!20}}cccc}
\rowcolor{red!20}
0 & 0 & \cellcolor{purple!45} 1 & ... & 0 & 1 \\
0& 1 & 0 & ... & 1 & 1 \\
1& 1 & 1 & ... & 0 & 1 \\
\vdots & & & & & \vdots\\
1& 0 & 0 & ... & 0 & 1 \\  
0& 1 & 0 & ... & 1 & 1  
\end{array}
\right)\right.
\end{equation*}
\caption{\label{subfig:a}}
\end{subfigure}
\end{minipage}
\begin{minipage}{0.33\columnwidth}
\vspace{-.2cm}\hspace{.4cm}
\captionsetup[subfigure]{oneside,margin={1.4cm,0cm},skip=-10pt}
\begin{subfigure}[b]{\columnwidth}
\input{img/SABM-SR_diagram.tikz}
\caption{\label{subfig:b}}
\end{subfigure}
\end{minipage}
\begin{minipage}{0.3\columnwidth}
\hspace{1cm}\vspace{.4cm}
\begin{subfigure}[b]{\columnwidth}
\centering
\includegraphics[width=0.55\columnwidth]{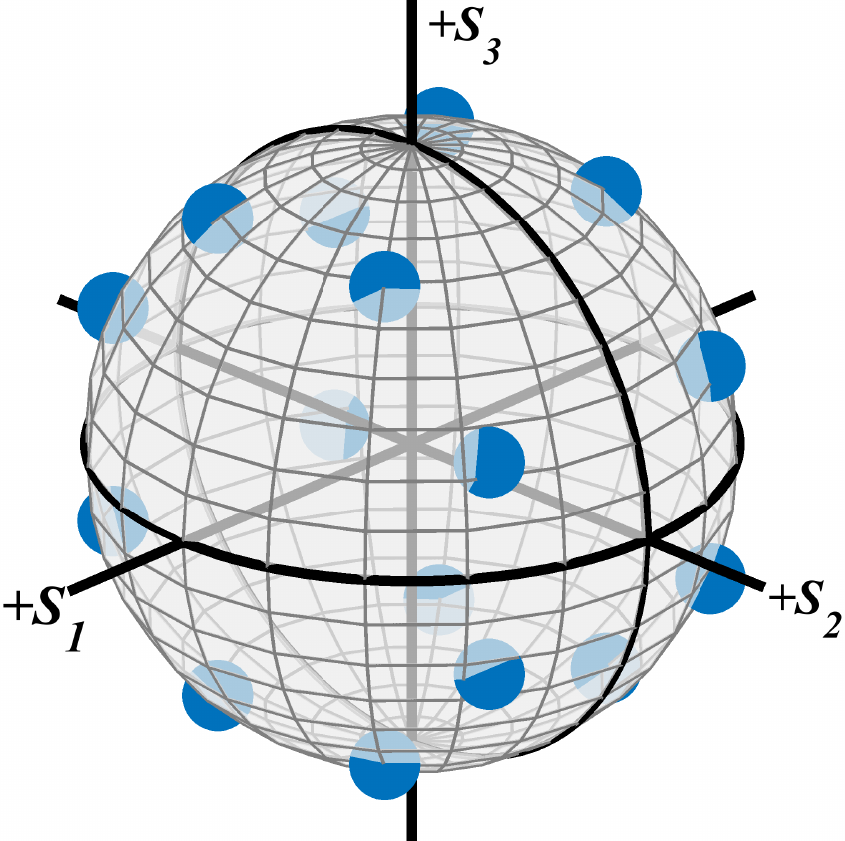}
\caption{\label{subfig:c}}
\end{subfigure}
\end{minipage}\vspace{-6mm}
\caption{Illustration of a PC array \subref{subfig:a}. Schematic diagram of one iteration of the SABM-SR algorithm \subref{subfig:b}, and representation of the 4D-64PRS modulation format on the Poincar\'{e} sphere \subref{subfig:c}.}
\label{fig:SABM-SR}
\end{figure*}
\begin{figure*}[!b]
\includegraphics[height=3.8cm,width=0.92\textwidth]{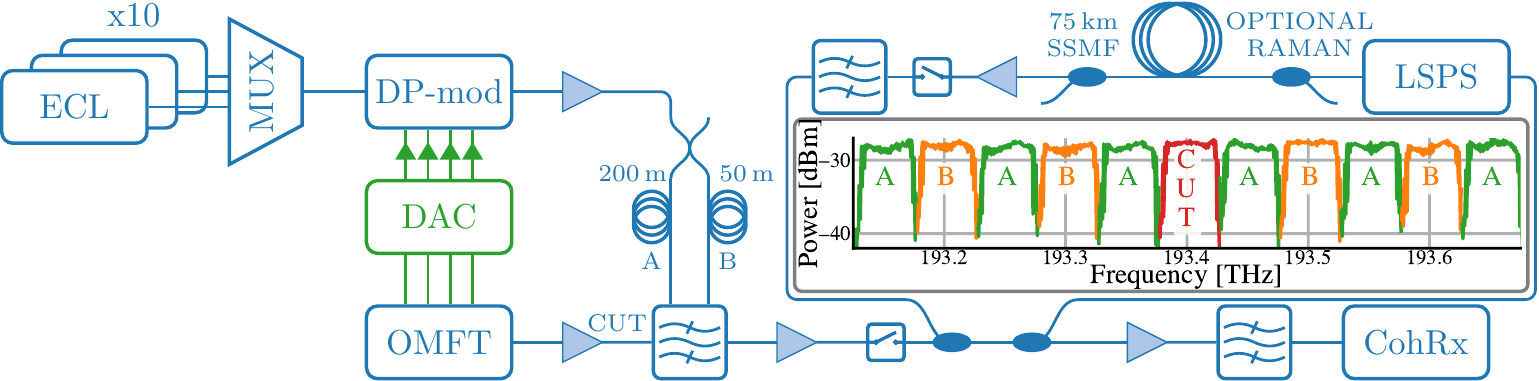}
\caption{Schematic diagram of the experimental testbed. On the right-hand side the received spectrum after $N=80$ circulations (6,000~km) of EDFA-only amplification. The CUT is depicted in the center position but is tested in all 11 positions in the experiment.}  
 \label{fig:setup}
\end{figure*}

In combination with \gls{FEC}, constellation shaping has been demonstrated to be a viable solution for providing additional signal-to-noise ratio (SNR) gains at a given \gls{SE}. In particular, geometrical shaping can be easily coupled with FEC and only requires straightforward modifications of the mapper and demapper. Recently, the \gls{4D-64PRS} format, introduced in \cite{BinChenJLT2019}, was demonstrated to outperform other notable 4D modulation formats (see e.g.,~\cite{Kojima2017JLT}) at a nominal \gls{SE} of 6 bit/4D-sym \cite{BinChenJLT2019, VanDerHeideOECC2019}, thus representing a viable solution for long-reach 400G (dual-carrier) transponders.  


In this work, we combine the low-complexity SABM-SR decoder and a PC-coded nonlinearity-tailored 4D-64PRS modulation format, enabling transmission of 11$\times$218 Gbit/s over transatlantic distances ($\geq$ 5,000 km) at 5.2 bit/symbol. Moreover, we demonstrate a total 30\% reach increase over \gls{PM-8QAM} and iBDD decoding. 

\section{Coded modulation with PC-coded 4D-64PRS and SABM-SR algorithm}
A PC code array (see Fig.~\ref{subfig:a}) consists of a $n\times n$ matrix where both rows and columns are allowed codewords in in a so-called \emph{component code} codebook. PCs are high performance HD codes thanks to the iBDD procedure. As illustrated in \ref{subfig:a}, such a procedure consists in iteratively decoding the received PC array using the algebraic BDD algorithm on the component code, first by rows and then by columns. The main limitations of the iBDD are \emph{failures} and \emph{miscorrections}. A failure occurs when the received codeword and any possible transmitted codeword differ by a number of bits greater than the error correction capability of the code $t$. 
On the other hand, when a codeword is found within $t$ bit positions but it is not the transmitted one, a miscorrection occurs. 

The idea of the SABM algorithm is to minimize both failures and miscorrections via \emph{bit marking} and \emph{bit flipping}. The bit marking is performed via a bitwise reliability measure on the received bits, such as the \gls{LLR}. The marking process consists in assigning each bit to one of following two classes: (i) highly-reliable bits (HRBs), (ii) highly-unreliable bits (HUBs). This is done by setting an optimal threshold on the magnitude of the \gls{LLR}. The HUB class is used to prevent miscorrections arising from the iBDD as the core HD decoder for product-like codes: whenever the BDD attempts to flip an input bit which is flagged as a HRB a miscorrection is detected. Failures are instead avoided by performing another decoding attempt after the least reliable bit of the incoming binary vector is flipped. A more comprehensive description of the SABM algorithm can be found in \cite{Yi2019}.

The SABM-SR algorithm extends SABM by updating the bit reliability using the iterative HD decoding \cite{Liga2019ECOC}. This is performed via SRs, which are a heuristic reliability metric proposed in \cite{Sheikh2019tcom}, and obtained by linearly combining the scaled BDD hard output bits with the channel reliability. The workflow of a single iteration of the SABM-SR algorithm is illustrated in Fig.~\ref{subfig:b} for PCs. For each output bit $\hat{b}_{i,j}$ in row $i$ and column $j$, the SABM row decoder output provides quantized information $u_{i,j}$ by assigning values $-1$ and $+1$ to successfully decoded bits 0 and 1, respectively. When a failure occurs $u_{i,j}$ is instead set to 0. The green blocks in Fig.~\ref{subfig:b} show how the SRs $\phi_{i,j}$ are calculated at each iteration for both row and column decoding, namely, $\phi_{i,j}=w u_{i,j}+l_{i,j}$, where $l_{i,j}$ are the channel LLRs and $w$ are optimized weights through density evolution \cite{Sheikh2019tcom} for row/column decoding. Based on the updated SRs, a new bit marking is performed and a new mask of HUBs $\psi_{i,j}$ is passed to the SABM column decoder which then repeats the process.

In this paper, SABM-SR is used in combination with the recently introduced 4D-64PRS modulation format operating at a nominal SE of 6 bit/sym (64 points in 4D) \cite{BinChenJLT2019}. The 4D-64PRS is obtained via a joint optimization of both constellation coordinates and binary labeling to maximize the \gls{GMI} in a nonlinear optical fiber channel. The resulting constellation is constant modulus in 4D and its representation in the Poincar$\acute{\text{e}}$ sphere is illustrated in Fig.~\ref{subfig:c}. Although the \gls{4D-64PRS} is specifically designed to minimize post SD-FEC \gls{BER} (via the GMI), because of the high rate of operation and its geometrical regularity, similar performance improvements are also expected with hybrid HD/SD decoders such as the SABM-SR. Moreover, for pure HD decoders pre-FEC \gls{BER} is a strong a performance predictor. Thus, 4D-64PRS is a convenient choice due to its improved pre-FEC \gls{BER} performance compared to other more conventional 6 bit/4D symbol formats such as PM-(star)-8QAM (as shown in Sec.~\ref{sec:results}). 
 
\section{Transmission Performance}\label{sec:results}
\begin{figure*}[!t]
\begin{minipage}{0.33\columnwidth}\vspace{-2.5mm}\captionsetup[subfigure]{margin={1cm,0cm},skip=-3pt}
\begin{subfigure}[b]{\columnwidth}
\includegraphics[width=\columnwidth]{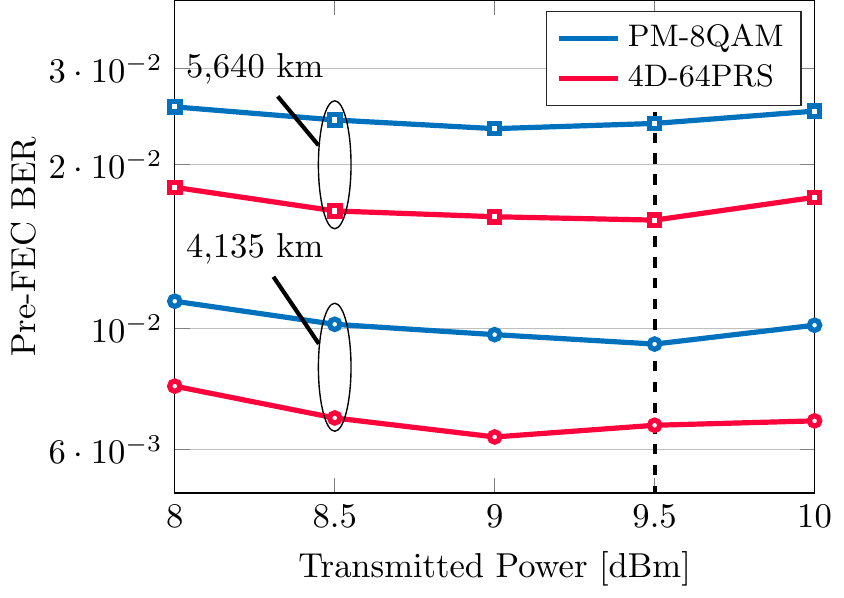}
\caption{\label{subfig:BERvsPow}}
\end{subfigure}\hfill\end{minipage}
\begin{minipage}{0.3\columnwidth}\hspace{2mm}\vspace{2.5mm}\captionsetup[subfigure]{margin={.6cm,0cm},skip=-1pt}
\begin{subfigure}[b]{\columnwidth}%
\includegraphics[width=\columnwidth]{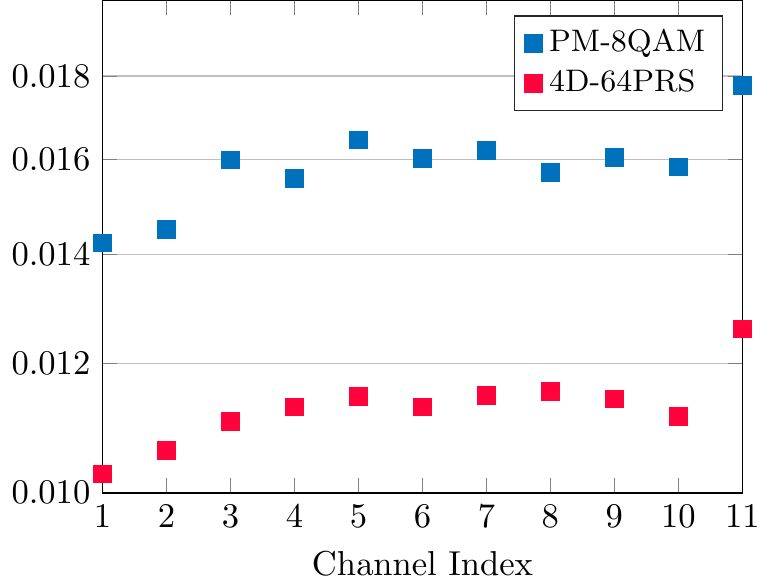}
\caption{\label{subfig:BERvsChIdx}}
\end{subfigure}
\end{minipage}\hfill
\begin{minipage}{0.33\columnwidth}
\captionsetup[subfigure]{margin={1cm,0cm},skip=-4pt}\vspace{-2.1mm}
\begin{subfigure}[b]{\columnwidth}
\includegraphics[width=\columnwidth]{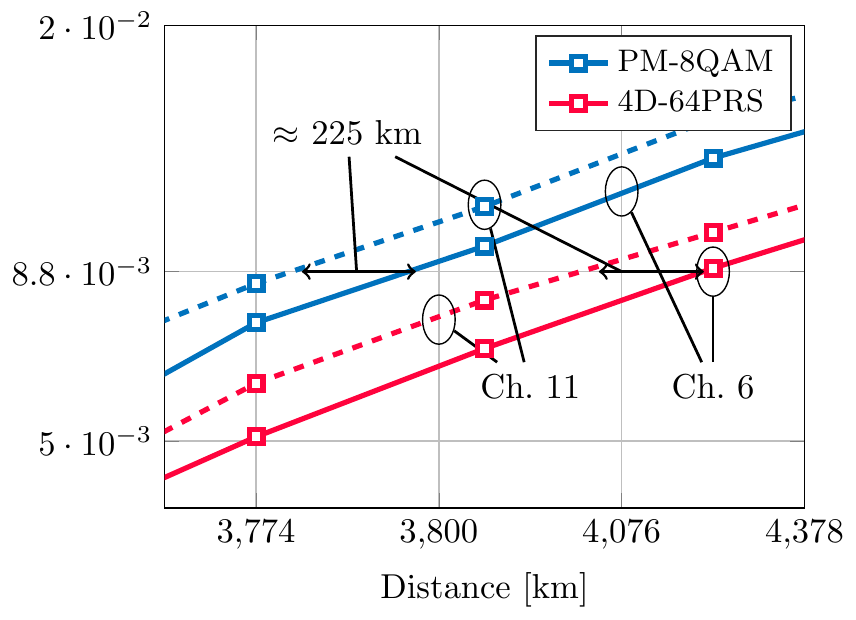}
\caption{\label{subfig:preFECvsDist}}
\end{subfigure}\end{minipage}\vspace{-4mm}
\caption{Experimental results. Pre-FEC BER vs total transmitted power for ch.~6 at $N=55$ and $N=75$ \subref{subfig:BERvsPow}. Pre-FEC BER vs.~channel index at 9.5 dBm total transmitted power and after $N=55$ circulations \subref{subfig:BERvsChIdx}. Post-FEC BER vs. transmission distance for ch.~6 at 9.5 dBm total transmitted power \subref{subfig:preFECvsDist}.}
\label{fig:exp_results}
\end{figure*}
The experimental testbed used for transmission is illustrated in Fig.~\ref{fig:setup}. Uncoded sequences of 2\textsuperscript{16} symbols are generated offline, filtered by a \glsentrylong{RRC} with 1\% roll-off at 41.79~GBd, pre-compensated for transmitter impairments, and uploaded to a 100-GSa/s \gls{DAC}. The \gls{CUT}, chosen among any of the 11 tested C-band channels, is modulated by the \gls{OMFT}, consisting of an \gls{ECL}, a dual-polarization optical modulator (DP-mod), and RF driving circuits, and subsequently amplified. The loading channels are generated using the multiplexed output of 10 \glspl{ECL} and the DP-mod. The channels are then amplified, split into even and odd, decorrelated by 10,200 (50~m) and 40,800 symbols (200~m) respectively, and finally multiplexed together with the \gls{CUT} on a 50-GHz grid using a multi-port \gls{OTF}. The resulting signal is transmitted over the recirculating loop, which consists of a \gls{LSPS}, a 75-km span of \gls{SSMF}, an \gls{EDFA}, an acousto-optical modulator, and an \gls{OTF} used for gain flattening. Fig.~\ref{fig:setup} (right) shows the optical spectrum after $N=80$ circulations, corresponding to a 6,000~km transmission. The optical signal exiting the loop is then optically amplified, filtered by a wavelength-selective switch, and digitized by a conventional intradyne coherent receiver. Offline digital signal processing includes front-end correction, frequency-offset compensation, chromatic dispersion compensation, and multiple-input multiple-output equalization with in-loop blind phase search.

To select the optimum transmitted power, a pre-FEC BER characterization was first performed. Two modulation formats were compared: Gray-labelled PM-(star)-8QAM and \gls{4D-64PRS}. Fig.~\ref{subfig:BERvsPow} shows the pre-FEC BER vs.~total transmitted power for the central of the 11 transmitted channels (ch.~6) after 4,135 km and 5,640 km ($N=55$ and 75, respectively). As the BER is essentially constant in the power range 8.5-9.5 dBm regardless of the distance and modulation format, a transmitted power of 9.5 dBm was selected as near-optimal for all distances and modulation formats investigated. Fig.~\ref{subfig:BERvsChIdx} shows the pre-FEC BER for the 11 transmitted channels over a distance of 4,135 km ($N=55$) and both modulation formats here investigated. The pre-FEC BER only exhibits minor fluctuations across the transmitted channels with ch.~11 being the worst-performing channel. However, for the purpose of demonstrating the SABM-SR gains and for experimental convenience, only the central channel (ch.~6) was used for evaluating the performance of the coded system. As shown in Fig.~\ref{subfig:preFECvsDist}, using ch. 11 as opposed to ch. 6 would result in a transmission distance penalty of about 225 km for both PM-8QAM, and 4D-64PRS modulation formats. Coding gains are, however, expected to be preserved for all transmitted channels.

For the evaluation of the coded transmission performance, the uncoded transmitted bits in the experimental traces were interleaved and scrambled to match the required number of randomly transmitted codewords. A PC using a 1-bit extended \gls{BCH} component code with information length, block length, and code rate of 239 and 256, and 0.87, respectively  is considered, resulting in a net SE of 5.2 bit/4D-sym. As discussed in \cite{Yi2019}, the SABM-SR algorithm requires both the optimization of the number of iterations $m$ over which miscorrection detection is performed and of the vector of weights $\boldsymbol{w}=[w_1,w_2,...,w_m]$, where $w_i$ represents the weight utilized at decoding iteration $i$ for both row and column decoding (see Fig.~\ref{subfig:a}). After numerical optimization, 
we found that $m=5$ and $\boldsymbol{w}=[3.42, 3.87, 4.08, 4.27, 4.49]$ are optimal values when a PC with extended \gls{BCH} (256, 239) as a component code is used.

\begin{figure}
    \centering
    \includegraphics[width=\columnwidth]{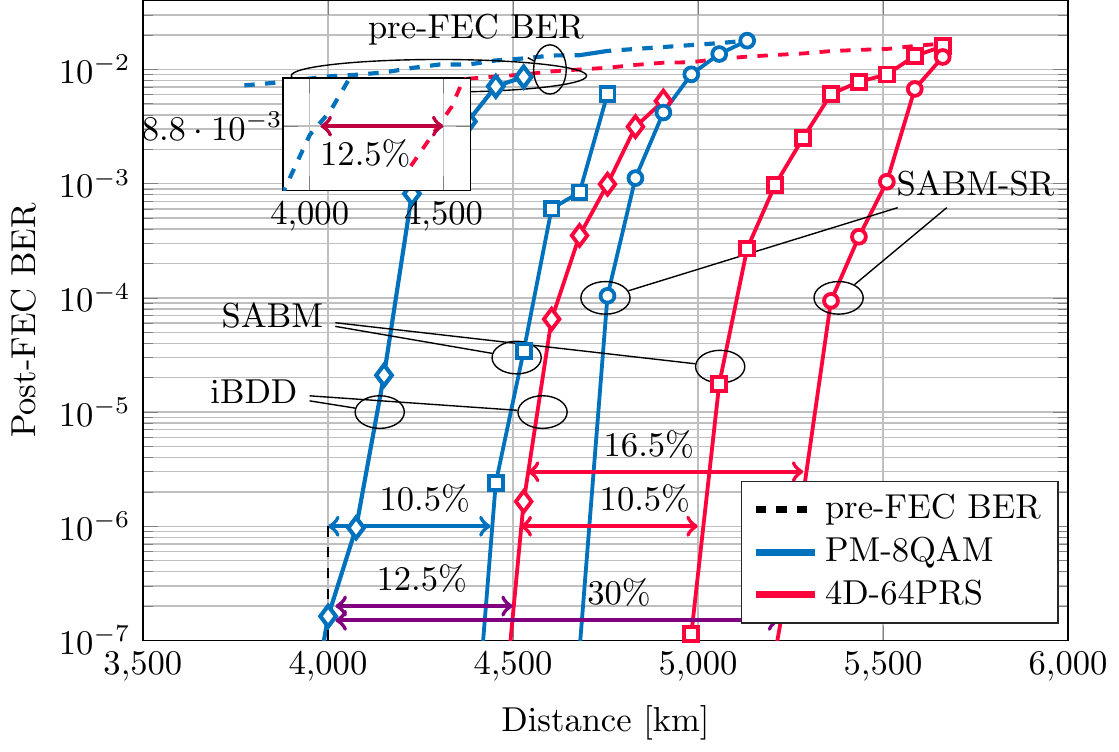}\vspace{-3mm}\caption{Performance of SABM (squares) and SABM-SR (circles) compared to a conventional iBDD decoder (diamonds) using both PM-8QAM and 4D-64PRS modulation formats. Arrows indicate gains at a post-FEC BER of $10^{-7}$.}
    \label{fig:SABM-SR_R=0.87}
\end{figure}

Fig.~\ref{fig:SABM-SR_R=0.87} shows the post-FEC \gls{BER} vs.~transmission reach for PM-8QAM and 4D-64PRS. When iBDD was used (diamonds), the 4D-64PRS (red) showed a 12.5\% reach increase at a post-FEC BER of $10^{-7}$ compared to PM-8QAM (blue). As expected, this gain is the same as the pre-FEC BER gain achieved by 4D-64PRS at BER=$8.8\cdot 10^{-3}$ (inset), corresponding to a post-FEC BER of $10^{-7}$ for iBDD. The SABM decoder (squares) yielded a 10.5\% reach increase when used with either 4D-64PRS or with PM-8QAM. Furthermore, SABM-SR achieved a 16.5\% reach increase compared to iBDD, regardless of the modulation format used. Thus, using SABM-SR led to an additional 6\% reach increase compared to SABM. Combining SABM-SR and 4D-64PRS (red circles curve) yielded a remarkable 30\% reach increase vs.~PM-8QAM with iBDD. Finally, Fig.~\ref{fig:SABM-SR_R=0.87} shows that transatlantic transmission ($\geq$ 5,000~km) was achieved using PC-coded 4D-64PRS and SABM-SR at a net SE of 5.2 bit/4D-sym (for BER$\leq 10^{-7}$). 

\vspace{-2mm}
\section{Comparison with other PC decoders}
In this section, the performance of SABM-SR is experimentally compared with two relevant baseline decoders: \emph{miscorrection-free} iBDD (MF-iBDD) and an SD turbo-product decoder (TPD) based on the Chase-Pyndiah algorithm \cite{Pyndiah1998}. MF-iBDD represents a lower-bound for the performance of iBDD, and is obtained preventing all miscorrection events in the decoding process via a genie-aided approach \cite{Hager2018}. The TPD is instead a \emph{fully-fledged} SD decoder for PCs with near-optimal decoding performance \cite{Pyndiah1998}.

\begin{figure}[!t]
    \centering
    \includegraphics[width=0.97\columnwidth]{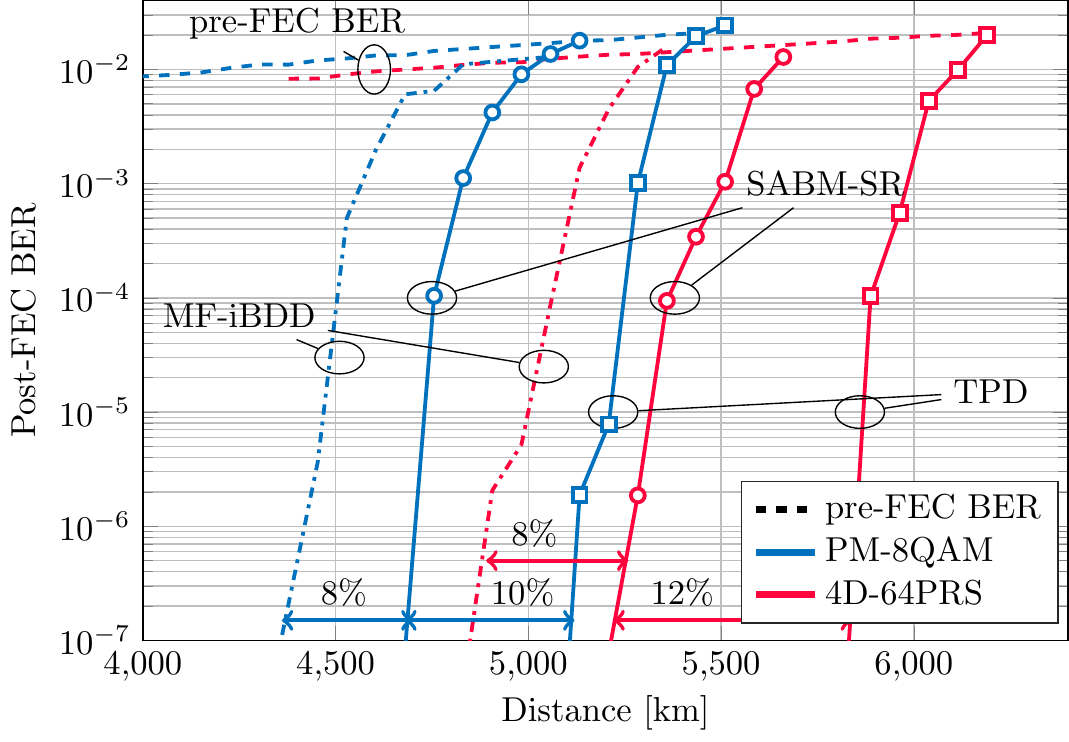}\vspace{-3mm}
    \caption{Performance of SABM-SR (circles) compared to MF-iBDD (dash-dotted lines) and TPD (squares) with both PM-8QAM and 4D-64PRS modulation formats. Arrows indicate gains at a post-FEC BER of $10^{-7}$.}
    \label{fig:TPD+MF_iBBD_R=0.87}
\end{figure}

Post-FEC BER performance vs. transmission distance is shown in Fig.~\ref{fig:TPD+MF_iBBD_R=0.87}. The SABM-SR decoder leads to an extension of the transmission reach of about 8\% compared to MF-iBDD in both the case of a PM-8QAM and 4D-64PRS transmission. We remark that such a reach extension is due to a twofold advantage of SABM-SR over MF-iBDD: i) the mitigation of decoding failures, which effectively extends the correction capability of the PC component code; ii) \emph{the miscorrection recovery} which consists not only in detecting a miscorrection, but also in some cases in correcting the errors leading to that miscorrection event. The TPD significantly outperforms SABM-SR (see Fig.~\ref{fig:TPD+MF_iBBD_R=0.87}), thanks to the full processing of the SD information. This enables an additional 12\% reach extension when 4D-64PRS is used. Similar coding gains (8\% and 10\% for MF-iBDD and TPD, respectively) are also observed for PM-8QAM. Although a detailed complexity analysis for SABM-SR is beyond the scope of this work, the TPD can be reasonably assumed to be within 1 and 2 orders of magnitude more complex than SABM-SR \cite[Sec.~IV-C]{Yi2019} or other recently proposed reliability-based iBDD algorithms  \cite{FougstedtOFC2019}.

\vspace{-2mm}
\section{Conclusions}
By combining a nonlinearity-tolerant 4D modulation format (4D-64PRS) and the SABM-SR decoding algorithm, we transmitted 11$\times$218 Gbit/s channels over a transatlantic distance at a net \gls{SE} of 5.2  bit/4D-sym. A 30\% reach increase vs. a conventional PM-8QAM transmission and HD decoding was also demonstrated. The results in this work highlight the potential of SABM-SR as a solution for low-power consumption/low-latency transceivers for long-haul dual-carrier 400G systems. 

\vspace{-2mm}
\section*{Acknowledgements}
\footnotesize{The work of~G. Liga, A. Sheikh and A. Alvarado has received funding from the European Research Council (ERC) under the European Union’s Horizon 2020 research and innovation programme (grant agreement No 757791). The work of A. Alvarado is supported by the NWO via the VIDI Grant ICONIC (project number 15685).~The work of B. Chen is partially supported by the National Natural Science Foundation of China (NSFC) under Grant 61701155. Partial funding from the Netherlands Organisation for Scientific Research (NWO) Gravitation Research Center for Integrated Nanophotonics (Under GA~024.002.033) is acknowledged.}

\vspace{-1.5mm}
\bibliographystyle{style/IEEEtran}
\bibliography{IEEEabrv,ref}
\end{document}

%% file: acronyms.tex
\newacronym{CUT}{CUT}{channel under test}

\newacronym{KK}{KK}{Kramers-Kronig}
\newacronym{CSPR}{CSPR}{carrier-to-signal power ratio}
\newacronym{KKRX}{KKRx}{Kramers-Kronig Receiver}
\newacronym{SSBI}{SSBI}{signal-signal beat interference}

\newacronym{GS}{GS}{geometric shaping}
\newacronym{SE}{SE}{spectral efficiency}
\newacronym{PS}{PS}{probabilistic shaping}
\newacronym{DSP}{DSP}{digital signal processing}
\newacronym{MIMO}{MIMO}{multiple-input multiple-output}
\newacronym{TDE}{TDE}{time domain equalizer}
\newacronym{FDE}{FDE}{frequency domain equalizer}
\newacronym{LMS}{LMS}{least means square}
\newacronym{DDLMS}{DD-LMS}{decision directed least means square}
\newacronym{FFE}{FFE}{feed-forward equalizer}
\newacronym{FBE}{FBE}{feedback equalizer}
\newacronym{BPS}{BPS}{blind phase search}

\newacronym{SMF}{SMF}{single-mode fiber}
\newacronym[plural=SSMFs]{SSMF}{SSMF}{standard single-mode fiber}
\newacronym[plural=FMFs]{FMF}{FMF}{few-mode fiber}
\newacronym{FMF12}{FMF12}{12 mode FMF}
\newacronym{MMF}{MMF}{multi-mode fiber}
\newacronym{SI}{SI}{step index}
\newacronym{GI}{GI}{graded index}
\newacronym{DCF}{DCF}{dispersion compensated fiber}

\newacronym{SDM}{SDM}{space division multiplexing}
\newacronym{MDM}{MDM}{mode division multiplexed}
\newacronym{WDM}{WDM}{wavelength division multiplexing}
\newacronym{DWDM}{DWDM}{dense wavelength division multiplexing}
\newacronym{LP}{LP}{linear polarized}
\newacronym[plural=MMUXs,firstplural=mode multiplexers]{MMUX}{MMUX}{mode multiplexer}
\newacronym{PL}{PL}{photonic lantern}
\newacronym{3DWG}{3DWG}{3D-waveguide}
\newacronym{MDL}{MDL}{mode dependent loss}
\newacronym{DGD}{DGD}{differential group delay}
\newacronym{DMGD}{DMGD}{differential mode group delay}
\newacronym{QSM}{QSM}{quasi-single-mode}
\newacronym{GIMMF}{GI-MMF}{graded-index multi-mode fiber}

\newacronym{SSB}{SSB}{single side band}
\newacronym{QPSK}{QPSK}{quadrature phase shift keying}
\newacronym{QAM}{QAM}{quadrature amplitude modulation}
\newacronym{PM-8QAM}{PM-8QAM}{polarization multiplexed 8-quadrature amplitude modulation}
\newacronym{RRC}{RRC}{root-raised-cosine}
\newacronym{4D-64PRS}{4D-64PRS}{
four-dimensional 64-ary polarization-ring-switching}
\newacronym{DP}{DP}{dual-polarization}
\newacronym[\glslongpluralkey=states-of-polarization]{SOP}{SOP}{state-of-polarization}
\newacronym{PM}{PM}{polarization-multiplexed}

\newacronym{ECL}{ECL}{external cavity laser}
\newacronym{CW}{CW}{continuous wave}
\newacronym[plural=DFBs]{DFB}{DFB}{distributed feedback laser}

\newacronym[plural=DACs]{DAC}{DAC}{digital-to-analog converter}
\newacronym{ADC}{ADC}{analog-to-digital converter}
\newacronym{PRBS}{PRBS}{pseudo-random bit sequence}
\newacronym{LO}{LO}{local oscillator}
\newacronym{EDFA}{EDFA}{erbium-doped fiber amplifier}
\newacronym{MZM}{MZM}{Mach-Zehnder modulator}
\newacronym{DP-MZM}{DP-MZM}{dual-polarization Mach-Zehnder modulator}
\newacronym{ChUT}{ChUT}{channel under test}
\newacronym{WSS}{WSS}{wavelength selective switch}
\newacronym[plural=VOAs]{VOA}{VOA}{variable optical attenuator}
\newacronym[plural=PDCRXs]{PDCRX}{PDCRX}{polarization diverse coherent receiver}
\newacronym{DSO}{DSO}{digital storage oscilloscope}
\newacronym{ASE}{ASE}{amplified spontaneous emission}
\newacronym{PBS}{PBS}{polarization beam splitter}
\newacronym{PD}{PD}{photodiode}
\newacronym{AOM}{AOM}{acousto-optical modulator}
\newacronym{BPD}{BPD}{balanced photo-diode}
\newacronym{OMFT}{OMFT}{optical-multi-format transmitter}
\newacronym{DPIQ}{DP-IQ-mod}{dual-polarization IQ-modulator}
\newacronym{ABC}{ABC}{automatic bias controller}
\newacronym{OTF}{OTF}{optical tunable filter}
\newacronym{LSPS}{LSPS}{loop-synchronous polarization scrambler}
\newacronym{OSA}{OSA}{optical spectrum analyzer}

\newacronym{OSNR}{OSNR}{optical signal to noise ratio}
\newacronym{BER}{BER}{bit error rate}
\newacronym{IL}{IL}{insertion loss}

\newacronym{SD}{SD}{soft-decision}
\newacronym{HD}{HD}{hard-decision}
\newacronym{SDFEC}{SD-FEC}{soft-decision forward error correction}
\newacronym{HDFEC}{HD-FEC}{hard-decision forward error correction}
\newacronym{FEC}{FEC}{forward error correction}
\newacronym{LDPC}{LDPC}{low-density parity-check code}
\newacronym{SABM}{SABM}{soft-aided bit marking}
\newacronym{SR}{SR}{scaled reliability}
\newacronym{PCs}{PCs}{product codes}
\newacronym{PC}{PC}{product code}
\newacronym{BCH}{BCH}{Bose-Chaudhuri-Hocquenghem}
\newacronym{iBDD}{iBDD}{iterative bounded distance decoder}
\newacronym{LLR}{LLR}{log-likelihood ratio}

\newacronym{AIR}{AIR}{achievable information rate}
\newacronym{AR}{AR}{achievable rates}
\newacronym{MI}{MI}{mutual information}
\newacronym{GMI}{GMI}{generalized mutual information}
\newacronym{BICM}{BICM}{bit-interleaved coded modulation}

\newacronym{OVNA}{OVNA}{optical vector network analyzer}
\newacronym{NIR}{NIR}{near infrared}
\newacronym{CD}{CD}{chromatic dispersion}
\newacronym{OTDR}{OTDR}{optical time domain reflectometry}
\newacronym{OFDR}{OFDR}{optical frequency domain reflectometry}
\newacronym{GPU}{GPU}{graphics processing unit}
\newacronym{SVD}{SVD}{singular value decomposition}
\newacronym{WGN}{WGN}{white Gaussian noise}
\newacronym{AWGN}{AWGN}{additive white Gaussian noise}
\newacronym{PDL}{PDL}{polarization dependent loss}
\newacronym{SPS}{sps}{samples-per-symbol}

%% file: img/SABM-SR_diagram.tikz
\resizebox{1.2\columnwidth}{!} 
{\begin{tikzpicture} 
\tikzset{fontscale/.style = {font=\tiny}}
\tikzstyle{block} = [draw, rectangle, minimum height=0.8cm, minimum width=2.3cm, rounded corners=0.1cm, align=center,font=\scriptsize] 
\node[block, minimum height=8.8em, minimum width=7.5em, fill=gray!10,dashed] (cont1) at (-28pt,29pt) {}; 
\node[block, minimum height=8.8em, minimum width=7.3em, fill=gray!10,dashed] (cont2) at (58pt,29pt) {}; 
\tikzstyle{Cir} = [draw, circle, minimum size=0.25em,font=\normalsize] 
\node[block,fill=red!20] (BDD1) at (-30pt,0) {SABM}; 
\node[block,fill=blue!20,right=0.7cm of BDD1,] (BDD2) {SABM}; 
\node[block, minimum height=2em, minimum width=2em, align=center, fill=yellow!20] (Mark1) at (40pt,45pt) {Bit \\ Marking}; 
\node[block, minimum height=1.5em, minimum width=2.5em, align=center, fill=yellow!20] (Mark2) at (-47pt,45pt) {Bit \\ Marking}; 
\node[block, minimum height=2em, minimum width=3.6em,fill=green!20] (SR1) at (-10pt,45pt) {};
\node[align=center, font=\scriptsize] (in) at (-86pt,25pt) {from previous \\ iteration};
\node[align=center, font=\scriptsize] (in) at (107pt,25pt) {to next \\ iteration};

\node[block, minimum height=2em, minimum width=3.6em,fill=green!20] (SR2) at (75pt,45pt) {}; 
\node[Cir,inner sep=0pt,fill=white] (mult) at (-19pt,45pt) {$\times$};
\node[Cir, right=0.25 cm of mult, inner sep=0pt,fill=white] (add) {$+$}; 
\draw[->] (BDD1.north-|mult)--++(0pt,10pt)node[right]{$\scaleto{u_{i,j}}{7pt}$}--(mult); 
\draw[->] (Mark2)--++(0pt,-24pt)node[right]{$\scaleto{\psi_{i,j}}{8pt}$}--(Mark2|-BDD1.north); 
\draw[->] (Mark1)--++(0pt,-24pt)node[right]{$\scaleto{\psi_{i,j}}{8pt}$}--(Mark1|-BDD2.north); 
\draw[->] (Mark2)--(Mark2|-BDD1.north); 
\node[above=0.33cm of add] (llr1) {$\scaleto{l_{i,j}}{9pt}$}; 
\node[above=0.35cm of mult] (w) {$\scaleto{w}{4pt}$}; 
\draw[->] (llr1)--(add); 
\draw[->] (w)--(mult); 
\draw[->] (mult)--(add); 
\draw[->] (BDD1.east)--++(13pt,0pt)node[above]{$\scaleto{\hat{b}_{i,j}}{10pt}$}--(BDD2); 
\draw[->] (add)--++(17pt,0pt)node[above]{$\scaleto{\phi_{i,j}}{8pt}$}--(Mark1); 
\node[Cir,inner sep=0pt,fill=white] (mult2) at (66pt,45pt) {$\times$}; 
\node[Cir, right=0.25 cm of mult2, inner sep=0pt,fill=white] (add2) {$+$}; 
\draw[->] (BDD2.north-|mult2)--++(0pt,10pt)node[right]{$\scaleto{u_{i,j}}{7pt}$} --(mult2); 
\node[above=0.31cm of add2] (llr2) {$\scaleto{l_{i,j}}{9pt}$}; 
\node[above=0.35cm of mult2] (w2) {$\scaleto{w}{4pt}$}; 
\draw[->] (llr2)--(add2); 
\draw[->] (w2)--(mult2); 
\draw[->] (Mark2)++(-25pt,0pt)--(Mark2); 
\draw[->] (BDD1)++(-40pt,0pt)--(BDD1); 
\draw[->] (mult2)--(add2); 
\draw[->] (add2)--++(15pt,0pt); 
\draw[->] (BDD2)--++(42pt,0pt); 
\node[] (end) at (80pt,0pt) {};
\node[align=center,font=\scriptsize] at (40pt,65pt) {Column \\ Decoder}; 
\node[align=center,font=\scriptsize] at (-50pt,65pt) {Row \\ Decoder}; 
\end{tikzpicture}}